\documentstyle[12pt]{article}

\newcommand{\doublespace}{
\newcommand{\mc}{\multicolumn}
\newcommand{\bce}{\begin{center}}
\newcommand{\ece}{\end{center}}
\newcommand{\be}{\begin{equation}}
    \renewcommand{\baselinestretch}{1.6}\large\normalsize}

\newcommand{\ee}{\end{equation}}
\newcommand{\bea}{\vspace{0.25cm}\begin{eqnarray}}
\newcommand{\eea}{\end{eqnarray}}

\def\PRA{{Phys. Rev.} A }

\doublespace

\begin{document} 

\centerline{\large \bf Experimental limit on Spontaneous Parametric Up Conversion}

\vskip 2cm

\centerline{ G.Brida, M.Genovese \footnote{ genovese@ien.it. Tel. 39 011 3919253, fax 39 011 3919259}, M. Gramegna, C.Novero \footnote{ Dedicated to the memory of our beloved friend and collaborator Carlo Novero.}}
\centerline{Istituto Elettrotecnico Nazionale Galileo Ferraris, Str. delle Cacce
91,I-10135 Torino.}

\vskip 1cm
{\bf Abstract}
\vskip 0.5cm 
We present a first experimental test of the existence of Spontaneus Parametric Up Conversion predicted in Ref. \cite{puc}.
The measurement has been made  using a ccd camera on the emission of a $LiIO_3$ crystal pumped by a 351 nm and/or a 789 nm lasers.
 We obtain an upper limit of 160 times on the ratio between intensity of Spontaneus Parametric Up Conversion and Spontaneous Parametric Down conversion.
\vskip 1.5cm

\vskip 2cm


\vspace{8mm}

Parametric Down Conversion  is a phenomenon where a pump laser converts into pairs of highly correlated photons fulfilling the condition, known as phase matching condition,  that the produced pair is such that the sum of frequencies of the produced photons is equal to the frequency of the pump photon and the vectorial sum of wave vectors  of the produced photons is equal to the wave vector of the pump photon.
A quantum theory of this effect \cite{Mandel} exists reproducing well all the results of the many experiments performed up to now.

The photon pair produced in PDC is an optical state with clear quantum properties and has often been used for tests of local realism \cite{Bell}. 

However, a classical description based on the use of Wigner function, which is positive for the states produced in PDC, is possible and has been developed recently in a series of papers \cite{santos,puc}, where PDC is treated as the interaction of a pumping laser wave and the zero-point vacuum waves inside a non-linear crystal.

The existence of such models, compatible with the nowadays experiments about Bell inequalities \cite{Bell}, shows that a classical local hidden variable model cannot be excluded completely yet. This is due to the well known detection loophole \cite{DL} that affects all the experiments about local realism performed up to now. This loophole derives from the fact that, because of low detection efficiencies, one needs a further additional hypothesis, stating that the observed sample of particles pairs is a faithful subsample of the whole \footnote{A recent experiment \cite{Win} performed using Be ions has reached very high detection efficiencies (around 98 \%), but in this case the two subsystems (the two ions) are not really separated systems during measurement and the test cannot be considered a real implementation of a detection loophole free test of Bell inequalities, even if it represents a relevant progress in this sense. }.

A very interesting point of the model of Ref. \cite{santos,puc} is that it presents various predictions which can be tested experimentally \cite{nos}.
In particular, an interesting prediction of this model, which is elaborated in Ref. \cite{puc}, is that a completely new phenomenon, dubbed Spontaneous Parametric Up Conversion, should exist. 

In synthesis, in the model of Ref. \cite{santos,puc} spontaneous PDC derives by the interaction inside a non-linear crystal of the high frequency $\omega _0$ (usually ultraviolet) pump laser with the zeropoint field in the vacuum (e.g. the one originating Casimir effect) which originates  two photons of lower frequencies ($\omega _2, \omega _3$) such that the equation
\be
\omega_0 = \omega_2 + \omega _3
\ee
for the frequencies and the equation 
\be
\vec{k}_0 = \vec{k}_2 + \vec{k} _3
\ee
for the wave vectors are satisfied. 

 If another laser of lower frequency $\omega _1$ is injected, with the right angle (see figure 1), a stimulated emission, with frequency $\omega _3$,  is produced satisfying the equations

\be
\omega_0  - \omega_1 = \omega _3
\ee

and

\be
\vec{k}_0 - \vec{k}_1 = \vec{k} _3
\ee

According to Ref. \cite{puc} a fluorescence is emitted also if the high frequency pump laser is turned off and only the second laser (the one with frequency $\omega _1$) remains. This phenomenon, due to the interaction of this laser with the zeropoint field (the vacuum),  has been dubbed Spontaneous Parametric Up Conversion and is a peculiar prediction of this model, which does not have equivalent in the traditional quantum optics description of SPDC. Therefore the search for this fluorescence represents a determinant test of the model.

The numerical calculations of Ref. \cite{puc} shows that  the intensity of this fluorescence should be a factor 2-3 smaller than PDC, but with a different angular distribution. If the incident laser (with frequency $\omega _1$) is normal to the non-linear crystal, the SPUC rainbow forms incomplete rings off-centre respect to the pump (see figure 3 of quant-ph 0203042 \cite{puc}), with frequencies ranging from $\omega_1$ and the UV. 
On the other hand, also in a configuration where a UV laser ($\omega _0$ in figure 1) was normally incident on a non-linear crystal originating a stimulated emission ($\omega _3$ in figure 1) by interacting with a second laser at a lower frequency ($\omega _1$ in figure 1), when the UV laser is turned off a SPDUC should appear. This emission should be distributed, because of phase matching conditions, on incomplete circles, of which the one with frequency $\omega _0$ passes through the point where the UV pump laser was and the one with frequency $\omega _3$ through the point where the stimulated emission was.
The remaining emission, with frequency varying from $\omega _0$ and $\omega _3$, lies between these two points. As in this case the region of emission is perfectly defined by the emission observed in stimulated PDC, this configuration is particularly well suited for an experimental search of SPUC.

In this paper we present a  first experimental test of this prediction, which gives severe constraints on the existence of SPUC.

More in details, we have pumped a Lithium iodate crystal, cut at the phase-matching angle of $51^o$, with an argon laser at 351 nm. A diode laser at 789 nm, with a power of 50 mW, was injected at the right angle for producing a stimulated emission (see figure 2).

Parametric down conversion was easily observed on a screen, even by naked eye, with a power of the argon laser lowered up to 50 mW (the maximum power of this argon laser is 500 mW).
If the IR laser was turned on an evident stimulated emission, at 633 nm, was visible (see figure 2), certifying that this second laser was entering the crystal with the right angle.
According to the prediction of Ref. \cite{puc,mar}, turning off the argon laser, a strong Parametric Up Conversion, produced by the interaction of the IR laser with zeropoint field, should be observed. In particular an emission with frequency ranging between $\omega _0$ (where the pump laser was) and $\omega _3$ (where the stimulated emission was) should be present. Thus the use of this experimental configuration allows an easy identification of the region where SPUC should be emitted and of the spatial distribution of frequencies. As the spectral response of our ccd camera, based on silicon sensitive elements, is centred in the visible region (with a quantum efficiency above 10 $\%$, in the whole visible region and near UVA and IR) an eventual SPUC emission should therefore be easily observed. Nevertheless, no signal of this emission appeared in our experiment.  

In order to obtain an upper limit to the ratio between SPUC and SPDC intensities, we have shot the SPDC with a ccd camera, which measures the time integral of the irradiance coming from the source. The image of SPDC was very evident even with a shot time as short as 1/10 of second and a 2.8 stop (the argon laser was eliminated by an anti-UV filter after the crystal). Then we turned off the argon laser and we searched for SPUC. No fluorescence appeared on the pictures even with long time exposures. In figure  3  we present a 16 seconds exposure (the longest time exposure available on our camera) on a ccd camera opened with a 2.8 stop when only the I.R. laser is injected into the non-linear crystal. Dark frame has been subtracted (even if this contribution is rather unimportant). Due to the long time of the exposure some diffused light of the laser into the crystal appears, but no SPUC is observed.

Considering that the response of a ccd camera is linear with the intensity of incident light (as we operate far from saturation regime), this means that the ratio between SPUC and SPDC intensities is lower than 160 times.  

In order to give a further confirmation of the absence of SPUC we have also taken some long time exposure on an ordinary camera film. In this case the response is non-linear, due to reciprocity effect, and thus no ratio between SPUC and SPDC can be derived.  However  rather a strong indication about the non-existence of SPUC can be obtained: in fact even with a 10 minutes exposure at 1.7 stop with a 1600 ISO film no signal of SPUC appears.    

With the purpose of performing a second test of the presence of SPUC, we have injected into the crystal a Nd-YAG laser at 1064 nm and a power of 0.51 W, when entering into the crystal, as well. Also in this case we have tuned the entrance angle in order to observe a well visible stimulated emission at 524 nm when both the UV and IR laser are turned on.  We have then repeated the same procedure as before: whilst PDC was observable even with a 1/30 of second exposure at 2.8 stop with the argon laser power tuned to 0.48 W, no SPUC appeared up to a 1/2 second exposure with the same stop and the 1064 nm laser at the same power. Unluckily in this case the high-power Nd-YAG laser strongly diffuses into the crystal and creates a strong background  for longer exposures: thus we were not able to derive more stringent limits on the ratio between SPUC and SPDC intensities in this case. 

Finally, also repeating the experiments with normal incidence of the laser no SPUC has been observed (with the same limits as before).

Altogether these experimental limits give a severe constraint on the model of Ref. \cite{santos,puc} and substantially excludes the very existence of Spontaneous Parametric Up Conversion.

In conclusion we have performed a first experimental test of existence of Spontaneous Parametric Up Conversion predicted in the model of Ref.   \cite{santos,puc}. The test was performed comparing the emission observed by a ccd camera when SPDC is present and  when SPUC is expected. No SPUC was observed up to a ratio of 160 times (with the 789 nm laser) respect to the intensity of SPDC produced at the same laser power.

\bigskip 

\vskip1cm \noindent {\bf Acknowledgements} \vskip0.3cm We would like to
acknowledge support of Italian Space Agency under contract LONO 500172.

\vfill \eject
{\bf Figure Captions}
\vskip 1cm

-Figure 1  Sketch of the source of PDC. NLC is a $LiIO_3$ crystal cut at the phase-matching angle of $51^o$. U.V. identifies the pumping radiation at 351 nm ($\omega _0$). The infrared beam (I.R.) at 789 nm ($\omega _1$) is generated by a diode laser. The dashed line identifies the stimulated radiation at 633 nm ($\omega _3$). 

 - Figure 2 Picture of the parametric down conversion produced by an argon laser injected into a $LiIO_3$ crystal. A second laser at 789 nm is injected as well.
The spot of this second laser and the stimulated emission at 633 nm appear clearly in the picture. The first one corresponds to the larger spot bleached because of saturation of ccd pixels. The second one corresponds to the spot on the other side respect to the centre of the emission. The concentric circles of different colours are the spontaneous emissions at different wavelengths. In absence of the 789 nm injected laser, exactly the same spontaneous emission appears, but of course the stimulated spot is missing. 
 
- Figure  3 Picture obtained with a 16 seconds exposure on a ccd camera with a 2.8 stop when only the I.R. laser is injected into the non-linear crystal. The field is the same of Figure 2. Due to long time of the exposure some diffused light of the laser into the crystal and thermal noise of ccd elements appear, but no SPUC is observed. The scale is the same of figure 2.

\end{document}